\documentclass[modern,twocolumn]{aastex63}
\begin{document}

\title{The Temporal Analysis of Light Pollution in Turkey using VIIRS data}
\shorttitle{Light pollution in Turkey}
\shortauthors{Yerli et al.}
    \received{receipt date}
    \revised{revision date}
    \accepted{acceptance date}
    \published{published date}
    \submitjournal{Ap. Sp.\&Sci.}

\author{S.K. Yerli}
    \affiliation{Department of Physics, Orta Doğu Teknik Üniversitesi, 06800, Ankara, Turkey.}
    \correspondingauthor{S.K. Yerli}
    \email{yerli@metu.edu.tr}
\author{N. Aksaker}
    \affiliation{Adana Organised Industrial Zones Vocational School of Technical Science, University of Çukurova, 01410, Adana, Turkey.}
    \affiliation{Space Science and Solar Energy Research and Application Center (UZAYMER), University of Çukurova, 01330, Adana, Turkey.}
\author{M. Bayazit}
    \affiliation{Remote Sensing and Geographic Information System, University of Cukurova, 01330, Adana, Turkey}
\author{Z. Kurt}
    \affiliation{Remote Sensing and Geographic Information System, University of Cukurova, 01330, Adana, Turkey}
    \affiliation{Space Science and Solar Energy Research and Application Center (UZAYMER), University of Çukurova, 01330, Adana, Turkey.}
\author{A. Aktay}
    \affiliation{Turkish State Meteorological Service, Regional Forecast Center, 01360, Adana, Turkey}
\author{M.A. Erdoğan}
    \affiliation{Landscape Architecture Department, Faculty of Architecture, Hatay Mustafa Kemal University, 31060 Hatay, Turkey.}

\begin{abstract}
Artificial Light pollution (AL) in Turkey and in Turkish observatories between 2012--2020 have been studied using the archival data of Visible Infrared Imaging Radiometer Suite (VIIRS) instrument.
The astroGIS database has been used in processing the data (\href{https://www.astrogis.org}{astrogis.org}) \cite{2020MNRAS.493.1204A}.
The total energy released to space from Turkey increased by 80\% in 2019 with respect to 2012.
In the span of the dataset, a steady and continuous increase has been observed throughout all cities of the country.
On the other hand, Dark Sky Park locations, East and Southeast Anatolian regions and mostly rural areas around the cities kept their AL level constant.
Four demographic parameters have been studied and they were found to be correlated very well with AL:
Population ($R \simeq 0.90$); GDP ($R \simeq 0.87$); Total Power Consumption ($R \simeq 0.66$) and Outdoor Lightening ($R \simeq 0.67$).
Contrary to countries acting to prevent AL increases, Turkey seems to be at the beginning of an era where AL will arithmetically increase throughout the country and enormous amount of energy will continuously escape to space and therefore will be wasted.
Therefore, a preventive legislation, especially for invaluable astronomical site locations such as TURAG, TUG, DAG and ÇAAM where each is counted as a truly dark site due to their SQM values, has to be enacted in Turkey, in very near future.
\end{abstract}


\section{Introduction}

Night sky is expected to be dark enough to conduct human based observations.
However, there is an ongoing human activity throughout on all parts of the Earth surface which gradually prevents humanity experiencing the dark night sky.
Thus, \textit{light pollution} can simply be defined as artificial light contributing to the night sky \citep{2020MNRAS.493.2463C, 2020AAS...23540103M, 2020ERCom...2....2S}.
The contribution is so large that one-third of the humanity cannot see and identify the Milky Way.
The simplest reason is actually due to the world population being accumulated in or around the cities \citep{falchi2016}.
Light pollution in the night sky makes observations of astronomical objects very difficult when cumulative light above the large cities degrade the quality of observations especially for the observatories \citep{Gronkowski2017}.
Light pollution is also an ecological problem besides its negative effect on astronomy \citep{Navara2007}.
Note also that, when the world-wide awareness considered, UNESCO has listed \textbf{the night sky} as a universal heritage\footnote{\href{https://astronomicalheritage.net}{astronomicalheritage.net}}.

The most recent review on AL observed from space and it is interpretation for the human activity on the surface is given in \cite{2020RSEnv.237k1443L}.
Light pollution is monitored and studied using many different measurement techniques:
 Sky Quality Meter (SQM) photometers \citep{Zamorano2016, Puschnig2019},
 Satellite base Defense Meteorological Satellite Program - Operational Line-Scan System (DMSP/OLS),
 International Space Station (ISS) nighttime light measurements \citep{Kuffer2018}
 and Visible Infrared Imaging Radiometer Suite (VIIRS) \citep{Levin2019}.

Through the modern human history, populations are found to be not evenly distributed over the surface area of Earth.
However, population always increases arithmetically and population density can be shown to correlate with other demographic parameters, especially human activities affecting the environment \citep{1999RSEnv..68...77E, hara2004, 2014RemS....6.1705S}, specific to this work ``the night sky'' \citep{falchi2016}.
Thus, it could easily be summarized that human activity is correlated with consumed and/or with wasted energy, therefore this energy, when it is observed from the space, can also be correlated with ``negative effect of human activity''.

The first studies on AL pollution in Turkey\footnote{\href{http://isikkirliligi.org}{isikkirliligi.org}} began under the leadership of TUBITAK National Observatory (TUG)\footnote{\href{https://tug.tubitak.gov.tr}{tug.tubitak.gov.tr}} which was initiated by Zeki Aslan, earlier director of TUG.
They have been using a simple photodiode based instrument, namely ``Sky Quality Meter - SQM'', to collect AL pollution values in the Zenith direction which can be carried out either personally or within a campaign to increase the awareness of \textit{dark skies}.
Since then, they have also been working on the legislation part of the awareness.
In 2005, they documented the results of both engineering and legislative studies on AL which led the group to apply for a change in the law (see details in \href{http://isikkirliligi.org}{isikkirliligi.org}).

Improper use of outdoor lighting has a negative impact on astronomical observatories in Turkey. 
An earlier study by \citet{aslan_2001} noted that the background brightness level has increased by 23, from 1986 to 1999 due to increase in investments in tourism on the Mediterranean coast and use of outdoor lighting for decorative purposes.
AL pollution has also been understood as an important issue in observatory site selection studies.
In their site selection studies \citet{aksaker2015} were made used DMSP/OLS (Defense Meteorological Satellite Program's Operational Linescan System) data (years 2012--2015) for AL pollution in Turkey.
Another national study is carried out by \citet{KOCSAN201339} and they noted AL pollution analysis for 2010 using the same satellite data but for the city Antalya only.
In Ege University Observatory, \citet{Devlen2018EgeG} reported that the sky brightness measurements in 2017 was shortened by 1.5 hours compared to 2010.
Impact of AL pollution is continued to be an important issue for observatories.

In this study, we aim to find Turkey's position in terms of energy released into the space.
For this purpose, the database introduced in \cite{2020MNRAS.493.1204A} has been adapted for Turkey.
The layering details of Artificial Light (AL) are published online in the astroGIS database
\footnote{\href{https://www.astrogis.org}{astrogis.org}}.
A subset of the dataset has been adapted from the database and is given in Fig. \ref{F:turkey}.

\section{GIS and Nighttime Dataset}
\label{sec:data}
Geographic Information Systems (GIS) and Remote Sensing, and their capability to capture, store, manipulate and display data, have found robust, easy to use and, time and cost efficient utility in the analyses of any spatial phenomena anywhere on/above/below the earth surface \citep{chang2009introduction}.
Advantages of GIS tools for spatial analysis along with progressively high precision and free-cost satellite-based remote-sensing datasets, become a key technology for environmental monitoring including human–environment interactions such as the economic, environmental, and social factors that influence settlement systems.

In this work, astroGIS database has been used \citep{2020MNRAS.493.1204A}.
In addition to GIS dataset, the demographic data for 2018 have been retrieved from the archival database of Turkish Statistical Institute\footnote{\href{https://tuik.gov.tr}{tuik.gov.tr}}.
Since there is no legislation enacted in the country, we aimed to find more correlations in the demographic dataset with the AL pollution.
Therefore, measured national power consumption parameters reported by the Turkish government in February 2020 has been retrieved and adopted to all cities are given in Table \ref{T:cities} \citep{EPDK2020}.

\begin{figure*}
\centering
\includegraphics[width=0.9\textwidth]{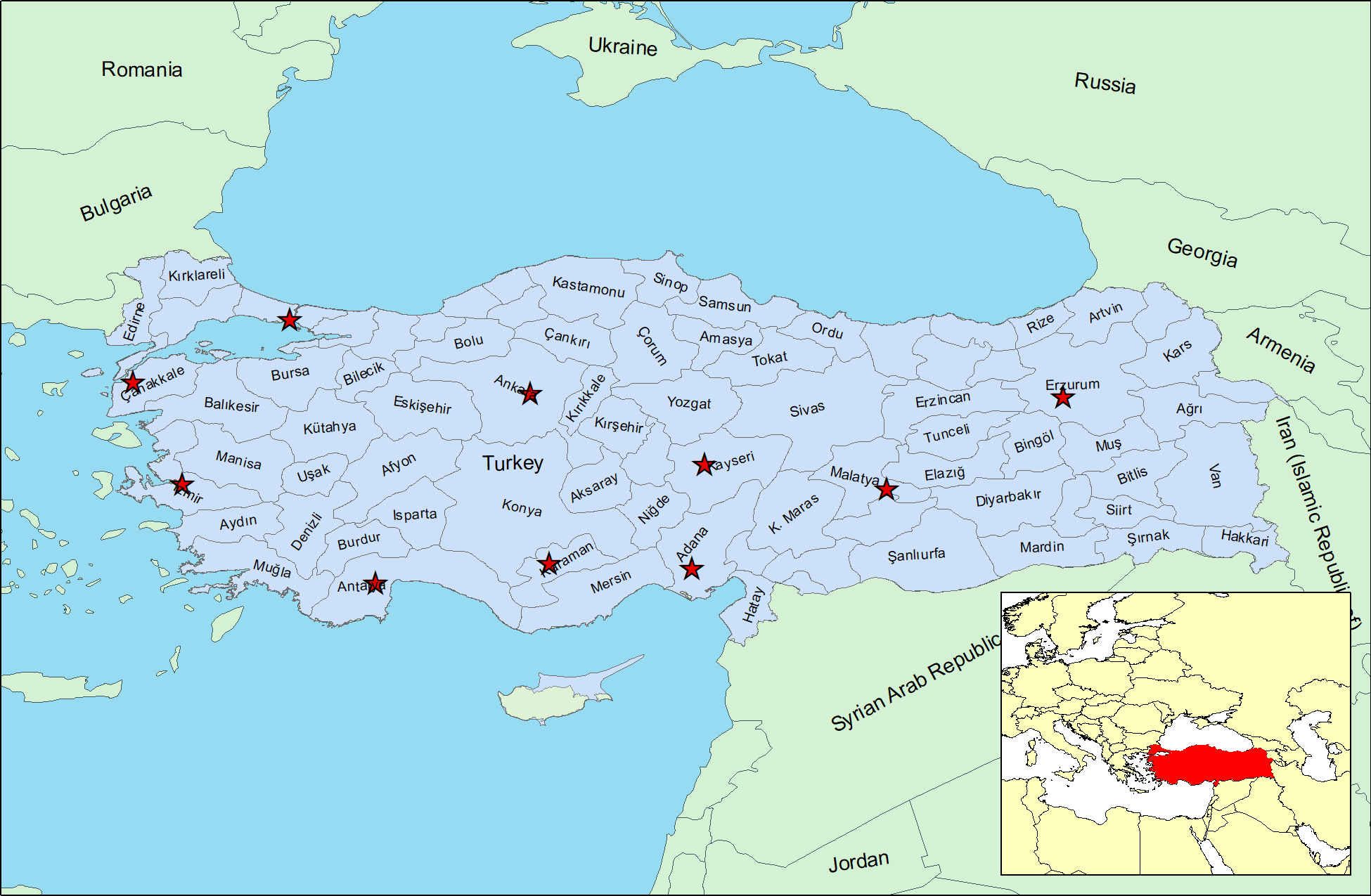}
\includegraphics[width=0.9\textwidth]{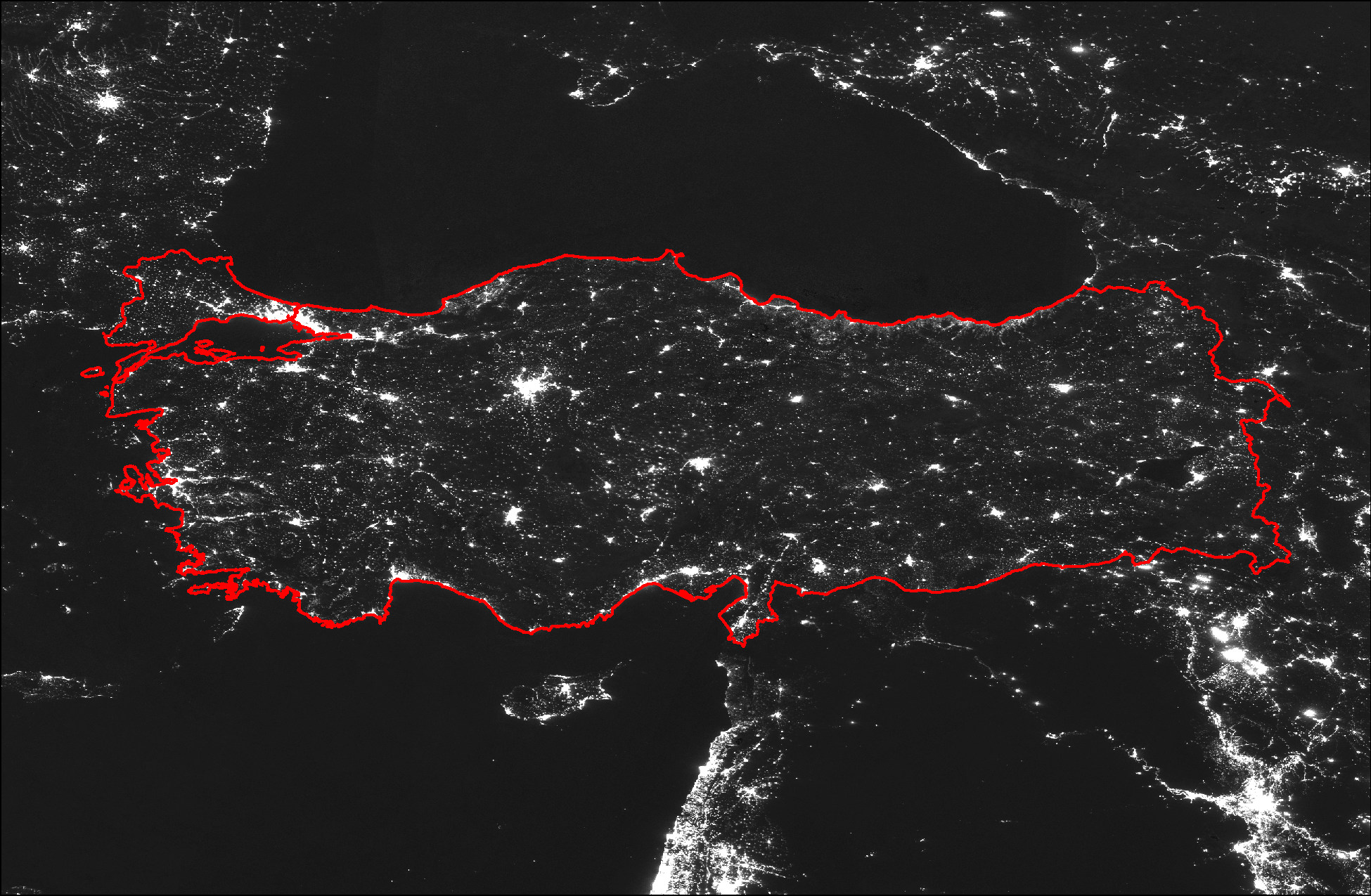}
\caption{%
Upper Panel: Demographic map of Turkey showing the name and boundary of each city.
Names not fitting to their boundaries are left as unnamed.
Both university owned and national observatories are marked with a red star.
Boundary of neighbouring countries are also drawn with no other details.
Lower Panel: Artificial Light (AL) distribution of Turkey for December 2019.
AL seen from space is colored as white.
As expected, İstanbul and other heavily populated major cities dominate the AL distribution.
Note also that geographically less populated regions, for example, rural areas, mountains, lakes etc. are colored with black.}
\label{F:turkey}
\end{figure*}

In constructing demographic dataset of Turkey the following information have been collected:
1) the country has 81 cities; 2) the city boundaries and total surface area were digitized from GADM dataset (see Table \ref{T:cities}).
It can be viewed in Fig. \ref{F:turkey} (upper panel).

The Visible Infrared Imaging Radiometer Suite (VIIRS) instrument on board SUOMI-NPP satellite in the Day Night Band (DNB) is used to acquire the nighttime data and they correspond to visible part of the spectrum.
Similar data analysis methods and techniques which were applied for France \citep{2020arXiv200604440A}, have been adapted in this work.
Overall view of the nighttime data for December 2019 is given as an example in the lower panel of Fig. \ref{F:turkey}.
The resultant spatial resolution of GEOTIFF images were 463 m per pixel.

\section{Analysis of the data}
\label{sec:analysis}
Using astroGIS database, a monthly averaged nighttime dataset has been produced.
This dataset contains 93 images.
Their dates range from April 2012 to December 2019.
Digitized GADM boundaries have been used, first to extract the surface area of Turkey.
Afterwards, each city has been extracted from the same dataset.
With respect to average light pollution value over the whole time span, above $3 \sigma$ values were excluded for each pixel using a pre-filtering algorithm written in house Python code.
In calculating pixel averages within each city boundary a model in Zonal Statistics tool of ArcGIS Desktop 10.4.1 has to be created to process 81 cities in total.
A city-based light pollution dataset is produced using monthly nighttime data for each city.
Earth Observation Group (EOG) updated VIIRS sensor calibration for Airglow \citep{2019ITGRS..57.9602U}. 
\cite{2020Sens...10.1964C} produced and published \citep{2019ITGRS..57.9602U} a mask for this new calibration.
This airglow correction has been applied to our dataset and used through out in all stages of data manipulation.

VIIRS launched in 2012, therefore, it spans a eight-years of AL data.
In this work, we aim to find AL variation over the full span of the satellite limiting to Turkey's surface area.

Accumulating this dataset for twelve month, yearly averaged data has been calculated and tabulated in Table \ref{T:result}.
A linear regression fit applied to yearly averages (column `L.R.') and it is given in the table along with goodness of fits (R$^2$).
Thus, possible variation in annual AL values could then be calculated (column $\Delta(\%)$).
Note that one of the main criteria in AL data is to locate and measure main AL contributing surface area or geographic locations throughout the country.
Therefore, luminous flux values of cities are given in Fig. \ref{F:dist}.

Overall view of variations were needed to visualize the effect of AL pollution through-out the country for the time span of the dataset emphasizing regional changes. 
This is given Fig. \ref{F:changes}.
 
\begin{figure*}
\centering
\includegraphics[width=\textwidth]{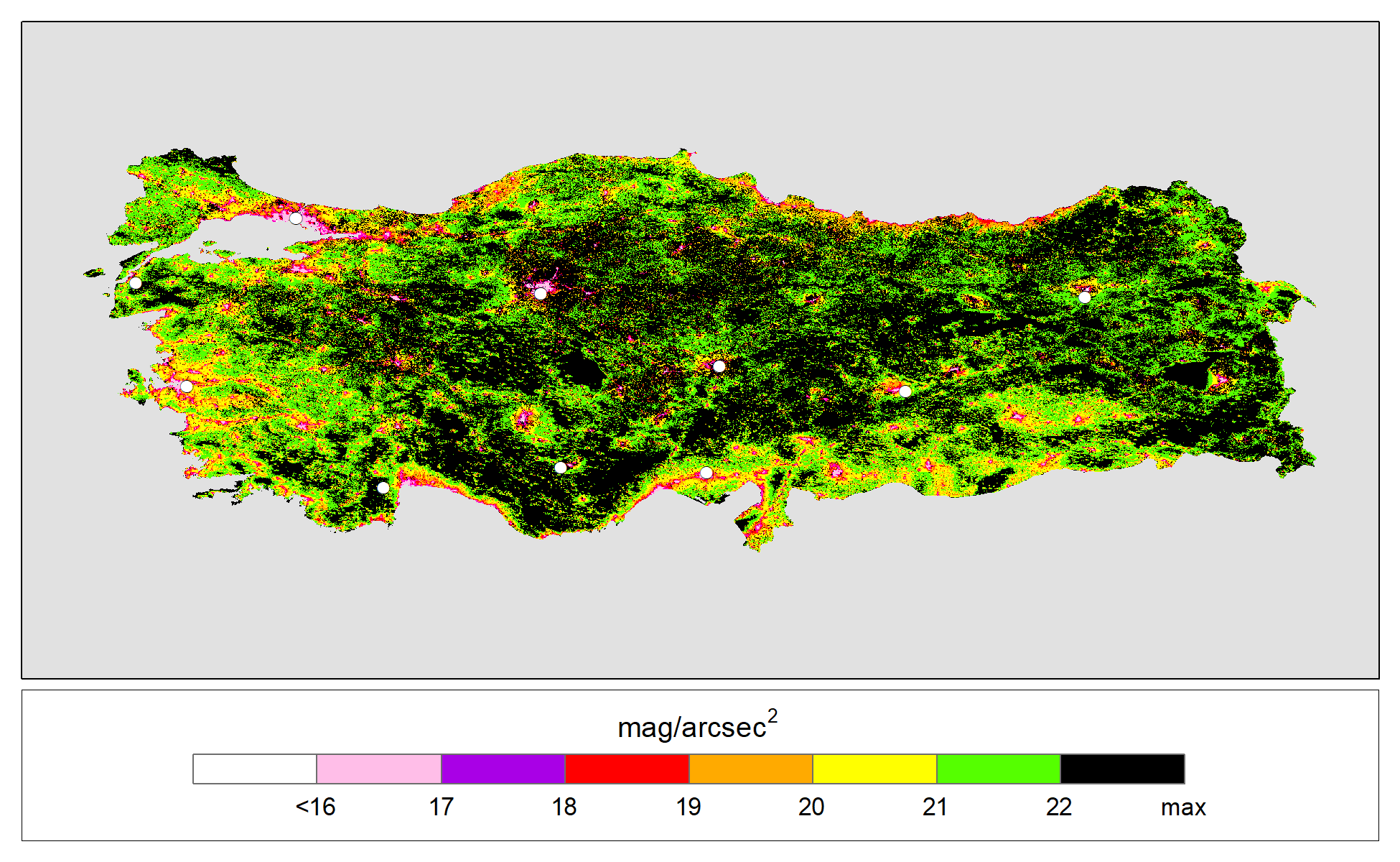}
\caption{%
Ten observatories of Turkey (white filled circles) overlaid on to the annual (2019) Artificial Light (AL) of the country.
The colors represents AL in unit of mag/arcsec$^{2}$.
Note that most of the observatories are severely effected from the AL.
However, dark regions (AL $>$ 22) still occupy a larger part of the country's surface area.%
}
\label{F:tr_obs}
\end{figure*}

\begin{figure*}
\centering
\includegraphics[width=0.95\textwidth]{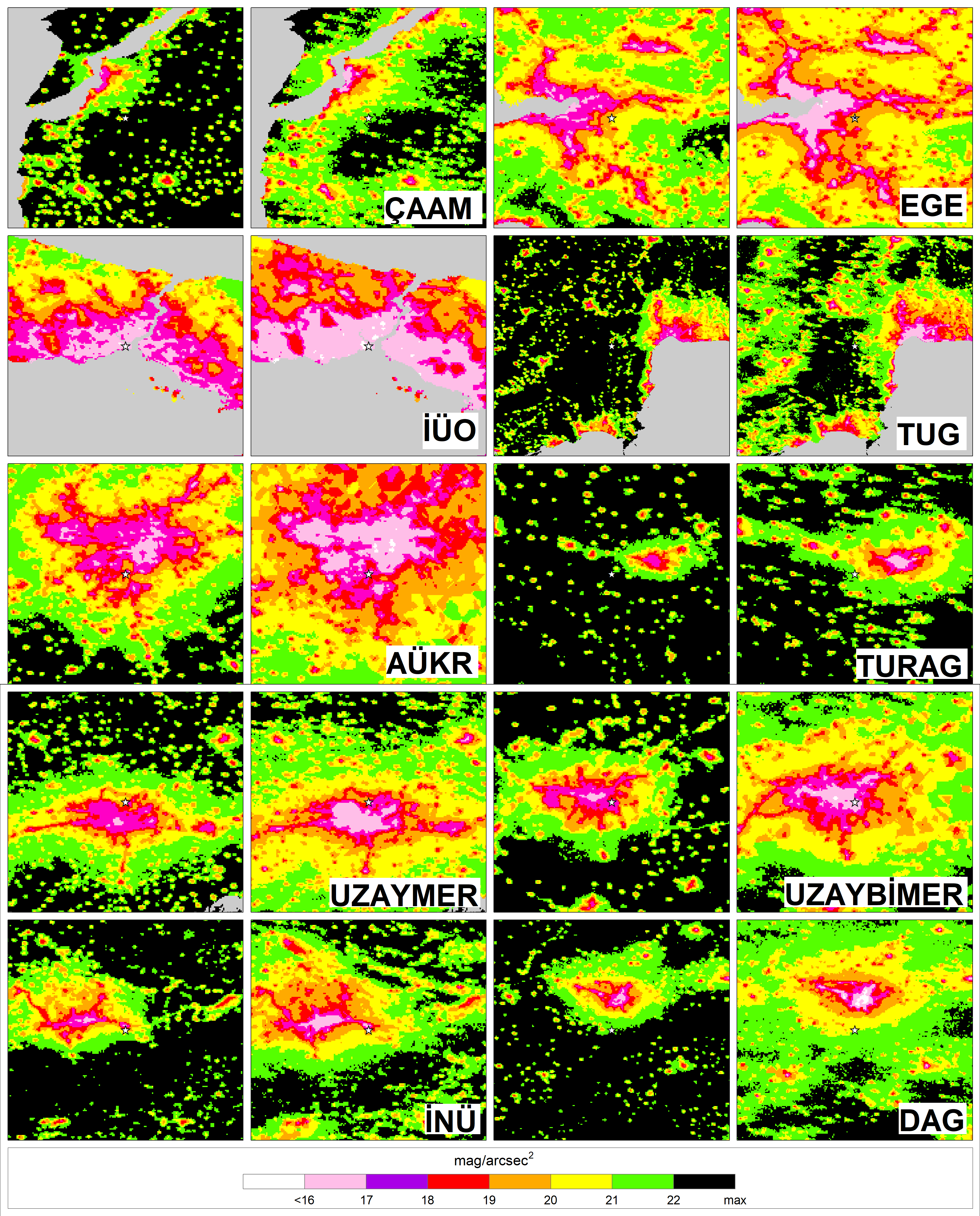}
\caption{%
Observatories of Turkey (white filled stars) overlaid on to the annual Artificial Light (AL) of the country for start and end of the dataset (left inset: 2012, right inset: 2019).
Observatories are listed according to longitude from left to right and north is up.
The scale is the same as Fig. \ref{F:tr_obs}.
AL for all observatories show the same trend as the country.%
}
\label{F:obs}
\end{figure*}

The Turkish observatory locations are given in Table \ref{T:obs}.
AL measurements were also carried for these locations (Table \ref{T:obs_result}).
They represent a single VIIRS pixel value which corresponds to 463$\times$463m and, therefore if it can be effected from nearby luminous pixels.
SQM values (in units of mag/arcsec$^2$ - mpsas) were calculated for each observatory location to have a good sense of site's astronomical quality.
Our AL dataset for each observatory is converted to SQM values using the following equation \citep{2020NatSR..10.7829S}:
\begin{equation}
20.0 - 1.9 \log(\textrm{AL})
\end{equation}
where AL is the VIIRS DNB value in nW cm$^{-2}$ sr$^{-1}$ units.
The contour map of SQM for the whole county (Fig. \ref{F:tr_obs}), and for all observatories (Fig. \ref{F:obs}) have also been produced to understand the effect of AL.
The SQM values of the observatories are given in Table \ref{T:obs_result}.

\section{Results and Discussions}
\label{sec:result}
\begin{figure}
\centering
\includegraphics[width=\columnwidth]{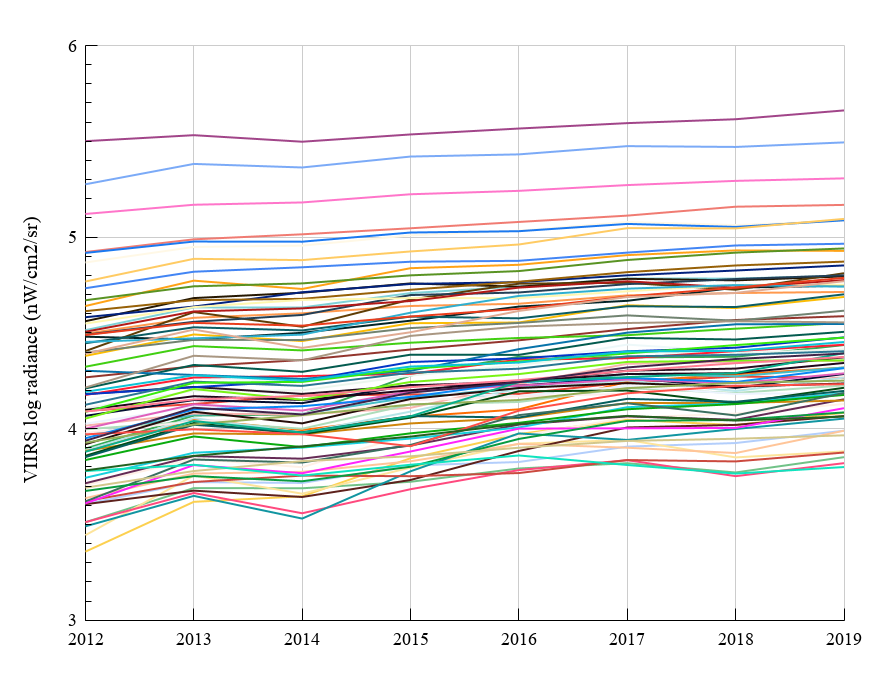}
\caption{%
	Artificial Light (AL) radiance averaged annually for all cities in between 2012--2019.
	Even though a general ``ascending'' trend exists in the graph, to be able to deduce a representative trend though out country, cities can easily be merged into two groups: over-luminous ($>$5) and luminous ($\le$5).
	Therefore AL of cities marked as ``luminous'' describes the general trend of the country: Variation of log values of AL in 2012: $\simeq$3.5--5.5 and in 2019: $\simeq$3.8--5.7.%
}
\label{F:dist}
\end{figure}

\begin{figure*}
\centering
\includegraphics[width=\textwidth]{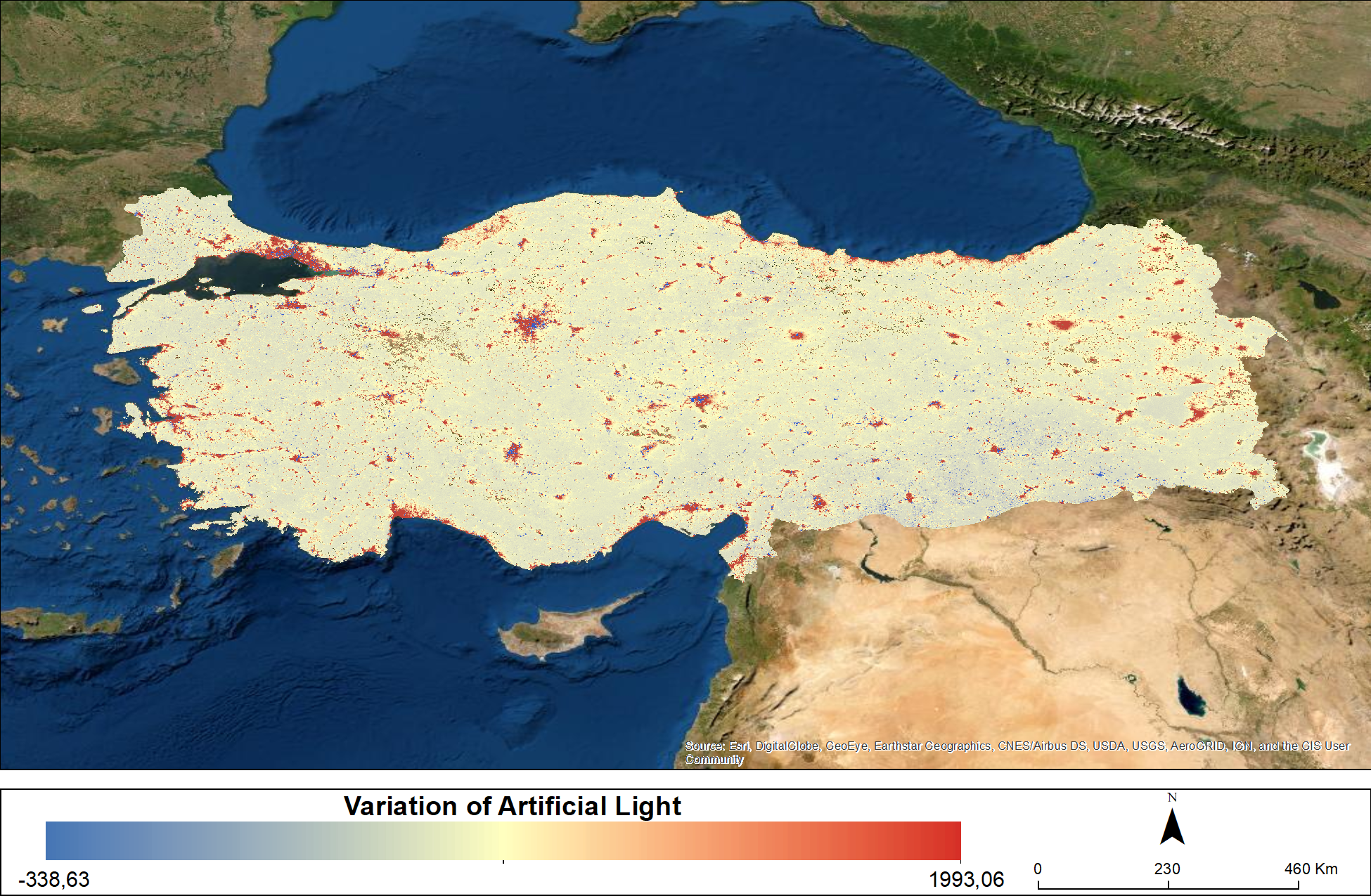}
\caption{%
Variation Map of Artificial Light (AL) in Turkey between 2012 and 2019.
AL's color gradient runs from blue (light pollution reduced) to red (light pollution increased).
The values are in W cm$^{-2}$ sr$^{-1}$.
See Section \ref{sec:analysis} for detailed explanation of the variation.%
}
\label{F:changes}
\end{figure*}

\begin{figure*}
\centering
\includegraphics[width=0.49\textwidth]{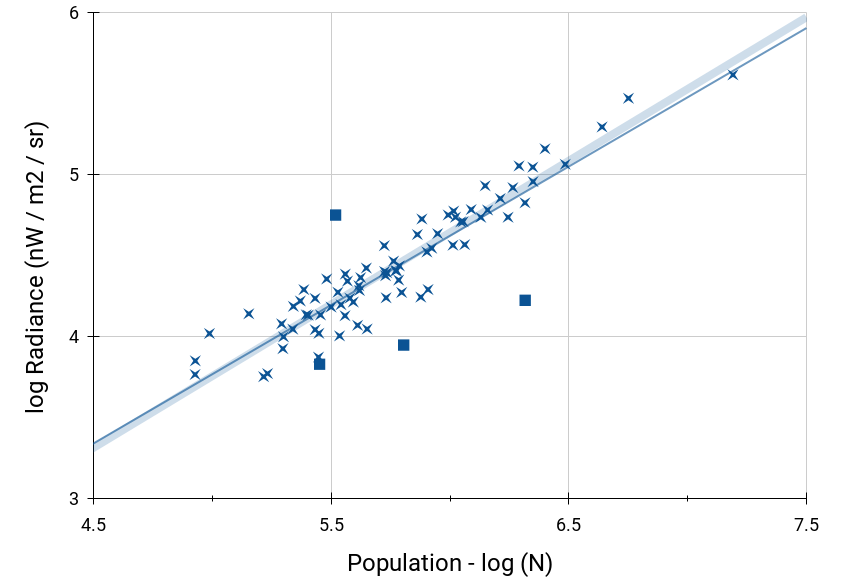}
\includegraphics[width=0.49\textwidth]{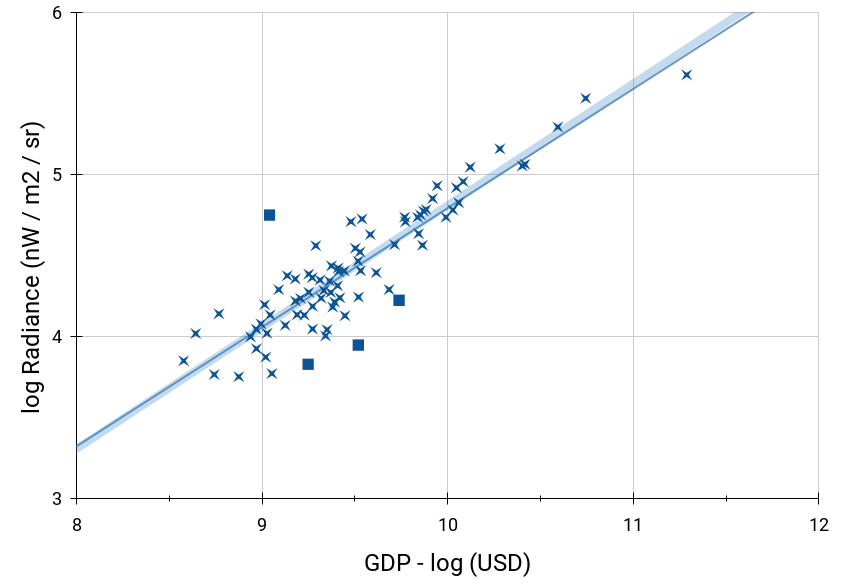}
\caption{%
Total radiance as measured by the satellite versus population (Left Panel) and GDP (Right Panel) for 2018 are plotted in logarithmic scales.
Thick solid lines represent the linear regressions for Population ($y=0.886x-0.674, R^2\sim 0.90$) and GDP ($y=0.752x-2.71, R^2\sim 0.87$).
Note that, both demographic values show good correlations with the radiance and when it is combined with Fig. \ref{F:dist} the country is continuously polluting the sky.
The cities marked with rectangles (Kilis, Siirt, Sivas and Şanlıurfa) became as outliers due to their AL values staying low or high (namely Siirt) with respect to their VIIRS pixels corresponding to their unit surface area.%
}
\label{F:pop_flux}
\end{figure*}

\begin{figure}
\centering
\includegraphics[width=\columnwidth]{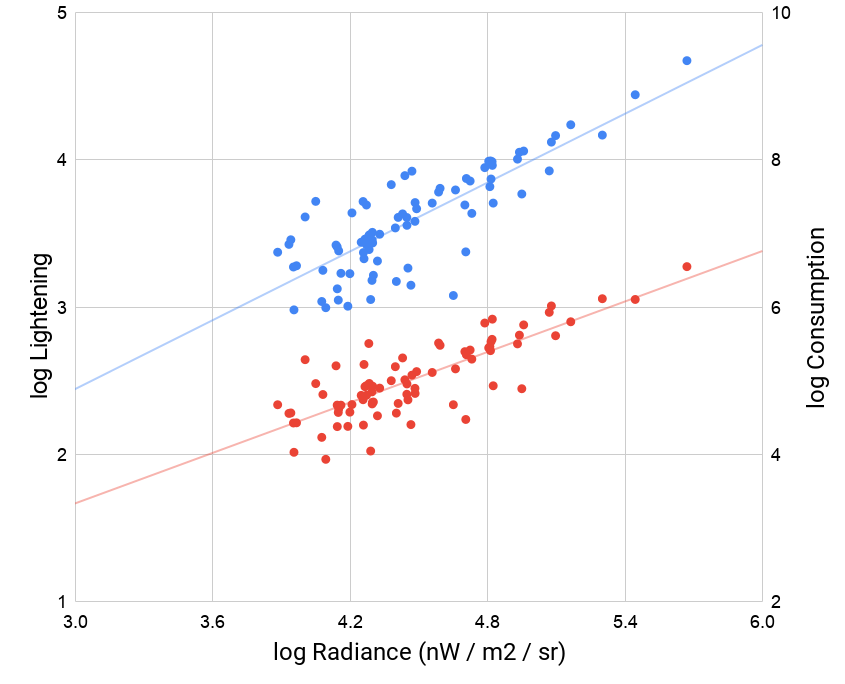}
\caption{%
Additional correlations between total radiance and two demographic values for each city are given with their linear fits for February 2019: Outdoor Lightening (left y-axis, in blue color) and Total Power Consumption (right y-axis, in blue color).
All three parameters are in logarithmic scale.
Thick solid lines represent the linear regressions for Power Consumption ($y=1.14x-0.09, R^2\sim 0.66$) and Lightening ($y=0.78x+0.11, R^2\sim 0.67$).
Lower part of the ``luminous'' cities ($<$ 4) show large variations around the fit, therefore decreases R$^2$.
However, the correlation for both values still exist with these outliers.%
}
\label{F:power}
\end{figure}

We investigated the light pollution dataset for Turkey.
The data runs from January 2012 to December 2019.
The whole dataset is created from our earlier astroGIS database \citep{2020MNRAS.493.1204A}.
The following outcomes have been noted after the analysis of the dataset:
\begin{itemize}
    \item
    The total energy released to space from Turkey increased by 80\% in 2019 compared to 2012.
    Energy release of Turkey during 2019 is 20\% higher than France \cite{2020arXiv200604440A}.
    
    \item İstanbul, as being the most populated city, produced 12\% of the total AL of the country during 2019.
    
    \item
    Even though there were disordered annual decreases in AL (Figure \ref{F:dist}), trend of AL for all the cities is positive and steady.
    
    Note that AL variation (Fig. \ref{F:changes}) over the whole country might give the impression of `no improvement'.
    In the reality of data, accumulation of light pollution over the large cities (an increase -- red colored pixels) might hide the improvements (a decrease -- blue colored pixels).
    
    \item
    During the time span of the dataset, all cities show a steady and positive increase in AL (Table \ref{T:result}).
    The maximum and minimum AL increase were observed in İstanbul and Tunceli, respectively.
    R$^2$ of the variation during 2012--2019 stays above 0.90 for 51 cities which proves that the country's AL continuously increases. 

    \item
    The following geographical ``points'', where they correspond to pixels in our dataset, Yusufeli/Artvin and Çorlu/Tekirdağ have the minimum (0.00) and maximum (883.6) values in VIIRS's luminous flux units, respectively. Note that these minimum and maximum values are for 2019 when the light pollution distribution over the country was considered.
    
    \item
    Turkish observatory locations have also been studied for AL and the following statistical outcomes have been observed:
    The brightest and the darkest observatory in 2019 are found to be İÖÜ (138.03) and TUG \& TURAG (0.10), respectively.
    The largest increase and decrease in AL, between 2012--2019, are found to be in UZAYBİMER and UZAYMER, respectively.
    Note that, AL for İÜO shows a decreasing trend which can be easily marked as an outlier since the observatory is within the most luminous city, İstanbul, showing a saturated trend in AL.

    We present AL contour maps in Fig. \ref{F:obs} covering 100 km surface area centered on all observatory locations.
    The AL values for observatories show the same trend as the country.
    We would like to emphasis the importance of reducing AL especially for Turkey's most important observatories, specifically for TUG (AL increase: 56\%) and DAG (AL increase: 40\%).
    It is not uncommon to predict the future with these outcomes that without having any controlling measures wealth of society will eventually degrade and destroy the astronomical night sky especially for these invaluable scientific investments on these observatories.
    Note also that, as it is given in Table \ref{T:obs_result}, for TURAG, TUG, DAG and ÇAAM, SQM values prove that these observatories are counted as truly dark sites.

    \item
    \cite{aksaker2015} found two main groups of astronomical observatory sites (17 in total).
    Group A includes the most suitable sites where all located in Southeast Anatolia Region.
    Even though their dataset for AL was 2012, changes observed throughout 2012--2019 (Figure \ref{F:changes}) show less light pollution (note dominated shades of blue) implying that the group A remained to be the most suitable site locations since all are in remote rural locations which are not effected very much from the human activities.

    \item
    Excluding two of most important astronomical parameters (elevation and cloud coverage), AL can also be used as a criteria to select specific locations, for example \textbf{Dark Sky Parks}.
    Turkey has dozens of such potential locations (see Fig. \ref{F:tr_obs}).
    The darkest cities among these locations are Kilis, Tunceli, Gümüşhane and Bayburt, once again confirming the main outcome: rural locations away from the human activity stayed as they are throughout 2012--2019.
    Note that the work with the full parameter set has been carried out and all suitable astronomical locations have been found in \cite{2020MNRAS.493.1204A}.

    \item
    Examples given in \cite{1999RSEnv..68...77E, hara2004, 2014RemS....6.1705S, 2020IJGI....9...32L} correlates human activity, say electric consumption, to the energy escaped to the space.
    Following this example as a base, our AL pollution dataset shows a strong correlation with both population (with 0.90 confidence) and GDP (with 0.87) (see Table \ref{T:cities} and Fig. \ref{F:pop_flux}).
    
    Even though there is a strong and obvious correlation between power consumption and AL observed in space, as it is noted in \citep{2014JQSRT.139..109S} before coming to an immediate conclusion, the observed flux has to be assessed locally (e.g. city by city) and corrected accordingly.
    The aim of this work, however, was focused on whether there exists a trend of AL or not; we confirm the trend in Fig. \ref{F:dist}.
    Due to free access to demographic values in Turkey, we have extended the impact of AL pollution by introducing two more consumption values \citep{1999RSEnv..68...77E, 2020IJGI....9...32L}.
    Observed good correlations (i.e. confidence level) for ``Total Power Consumption'' and ``Outdoor Lightening'' are 0.66 and 0.67, respectively (see Fig. \ref{F:power}).
    
    \item
    Contrary to countries acting to prevent AL increases, Turkey, with the values presented in this work, seems to be at the beginning of an era where AL will arithmetically increase throughout the country and enormous amount of energy will continuously escape to space and therefore will be wasted.
    Therefore, to overcome expected worst scenario where other countries have been faced with in past decades (such as France, see \citealp{2020arXiv200604440A}) a preventive legislation has to be enacted in Turkey, in the very near future.
\end{itemize}

\section*{Acknowledgements}
This research was supported by the Scientific and Technological Research Council of Turkey (TÜBİTAK) through project number 117F309. This research was also supported by the Çukurova University Research Fund through project number FYL-2019-11770.

\paragraph{Compliance with ethical standards} The authors declare that they have no potential conflict and will abide by the ethical standards of this journal.

\bibliographystyle{aasjournal}
\bibliography{T3-main}

\begin{table*}
    \caption{%
    Five demographic values for all cities in Turkey.
    Area is in square km.
    GDP is in billions US dollars and population is for 2018.
    GDP data is taken from TÜİK in units of thousands of Turkish Lira and an exchange rate of December 31st, 2019 is applied to obtain GDP in US dollars.
    L$_{out}$ and $\Sigma P_{cons}$ represent the values of \textit{Outdoor Lightening} and \textit{Total Power Consumption} for February 2020, respectively.
    See section \ref{sec:data} for the discussion.%
    }
    \centering
    \begin{small}
    \begin{tabular}{@{}%
        l@{~~}r@{~~}r@{~~}r@{~~}r@{~~}r@{\hspace*{0.5em}}
        l@{~~}r@{~~}r@{~~}r@{~~}r@{~~}r@{}%
        }
    \hline 
    City   & Area & Pop. & GDP & L$_{out}$ & $\Sigma P_{cons}$ &
    City   & Area & Pop. & GDP & L$_{out}$ & $\Sigma P_{cons}$ \\
    \hline
Adana		&15843	&2237940	& 12.1 &11469	&572566		&K.Maraş	&14586	&2232374	&  5.2 &6401	&302219	\\ 
Adıyaman	&7455	&626465		&  2.3 &3088	&91657		&Karabük	&3314	&579257		&  1.5 &2560	&46580	\\ 
Afyon		&14752	&729483		&  3.8 &6241	&144753		&Karaman	&8334	&800165		&  1.7 &1695	&46483	\\ 
Ağrı		&10491	&536199		&  1.4 &1492	&36262		&Kars		&10529	&1440611	&  1.1 &1407	&25332	\\ 
Aksaray		&7626	&337800		&  2.2 &2755	&63123		&Kastamonu	&13459	&1154102	&  2.1 &4931	&85427	\\ 
Amasya		&6216	&5639076	&  1.8 &2858	&48362		&Kayseri	&15953	&838778		&  8.8 &10106	&316369	\\ 
Ankara		&25731	&2511700	& 55.4 &27665	&1268170	&Kilis		&1428	&983142		&  1.8 &2357	&47128	\\ 
Antalya		&21184	&170875		& 19.1 &17297	&631273		&Kırıkkale	&4671	&408809		&  2.8 &2636	&159177	\\ 
Ardahan		&4863	&1110972	&  0.4 &1125	&11078		&Kırklareli	&6387	&303010		&  1.2 &2054	&33347	\\ 
Artvin		&7868	&1228620	&  1.1 &2864	&36451		&Kırşehir	&6780	&362861		&  0.6 &1115	&37032	\\ 
Aydın		&8038	&219427		&  5.9 &7465	&225340		&Kocaeli	&4044	&754198		& 25.2 &8410	&848109	\\ 
Balıkesir	&14773	&279812		&  7.7 &9778	&278824		&Konya		&40893	&343212		& 13.2 &14608	&407950	\\ 
Bartın		&2583	&348115		&  0.9 &2665	&35955		&Kütahya	&12217	&1029650	&  3.3 &4651	&132832	\\ 
Batman		&4401	&316126		&  2.1 &1517	&70481		&Malatya	&11544	&1348542	&  3.4 &5082	&129279	\\ 
Bayburt		&3421	&270796		&  0.4 &990	&8551		&Manisa		&13267	&330280		& 10.7 &7407	&339163	\\ 
Bilecik		&4649	&3056120	&  1.9 &2129	&166094		&Mardin		&10659	&218243		&  3.2 &1839	&54939	\\ 
Bingöl		&8217	&542157		&  1.0 &1330	&23799		&Mersin		&15213	&638956		& 11.2 &11253	&415279	\\ 
Bitlis		&8805	&195789		&  1.0 &2346	&24932		&Muğla		&12875	&1055412	&  7.2 &7177	&260722	\\ 
Bolu		&8266	&530864		&  2.4 &2458	&89022		&Muş		&8243	&612747		&  1.3 &1907	&26778	\\ 
Burdur		&7261	&1037208	&  1.6 &2648	&63407		&Nevşehir	&5137	&808974		&  1.5 &3216	&50482	\\ 
Bursa		&10886	&1756353	& 26.1 &13188	&1036221	&Niğde		&7187	&84660		&  1.8 &3124	&79238	\\ 
Çanakkale	&9838	&413903		&  4.1 &4293	&203743		&Ordu		&5929	&2073614	&  3.3 &6782	&99933	\\ 
Çankırı		&7915	&591098		&  1.0 &2407	&41435		&Osmaniye	&3077	&370509		&  2.6 &3037	&319829	\\ 
Çorum		&12309	&234747		&  2.6 &4057	&65379		&Rize		&3678	&1136757	&  2.2 &4359	&47403	\\ 
Denizli		&10915	&762062		&  7.4 &6572	&290005		&Sakarya	&4701	&421200		&  7.3 &6038	&324691	\\ 
Diyarbakır	&14952	&887475		&  5.9 &4327	&196251		&Samsun		&9529	&596053		&  6.9 &9821	&255865	\\ 
Düzce		&2388	&2069364	&  2.5 &2891	&82948		&Şanlıurfa	&19750	&283017		&  5.4 &1686	&37383	\\ 
Edirne		&6166	&448400		&  2.5 &2725	&84080		&Siirt		&5059	&416367		&  1.1 &2370	&29735	\\ 
Elazığ		&9350	&164521		&  2.8 &3590	&90339		&Sinop		&6010	&84843		&  0.9 &5221	&91340	\\ 
Erzincan	&11616	&280991		&  1.5 &1643	&51315		&Sivas		&28427	&253279		&  3.3 &4091	&192925	\\ 
Erzurum		&24656	&1628894	&  3.5 &5859	&77754		&Şırnak		&8018	&608659		&  1.9 &1198	&47164	\\ 
Eskişehir	&14058	&444914		&  7.0 &4935	&247879		&Tekirdağ	&6641	&529615		&  9.8 &8848	&606927	\\ 
Gaziantep	&7172	&1840425	& 11.5 &9168	&684888		&Tokat		&9853	&198249		&  2.4 &5124	&67142	\\ 
Giresun		&6841	&15519267	&  1.9 &5209	&55142		&Trabzon	&5069	&97319		&  4.8 &8372	&118941	\\ 
Gümüşhane	&6658	&4367251	&  0.7 &1871	&26633		&Tunceli	&8105	&199442		&  0.6 &957	&10649	\\ 
Hakkari		&6331	&285410		&  1.1 &1013	&23949		&Uşak		&4927	&270976		&  2.3 &3447	&154947	\\ 
Hatay		&5794	&379405		&  8.3 &9708	&364341		&Van		&21545	&248458		&  3.0 &5082	&85245	\\ 
Iğdır		&4012	&1407409	&  0.9 &1091	&16992		&Yalova		&651	&142490		&  2.2 &1776	&64922	\\ 
Isparta		&8879	&361836		&  2.6 &3821	&78671		&Yozgat		&13160	&538759		&  1.9 &4063	&49076	\\ 
İstanbul	&4767	&242938		&194.2 &47129	&3540493	&Zonguldak	&3058	&392166		&  3.4 &7807	&102832	\\ 
İzmir		&12435	&1953035	& 39.3 &14716	&1298341	&		&	&		&      &	&	\\ 
    \hline 
    \end{tabular}
    \end{small}
\label{T:cities}
\end{table*}
\begin{table*}
    \caption{%
    Annual Artificial Light (AL) for all cities of Turkey.
    Light pollution values are in units of in nW cm$^{-2}$ sr$^{-1}$.
    Two different values are given for AL: Ave-All, Ave-19 representing average of annual AL values in between 2012--2019, for 2019, respectively.
    L.R. and $R^2$ columns are the slope of linear regression and its correlation coefficient of the regression, respectively for annual AL values in between whole data range.
    See section \ref{sec:data} for the discussion on the trend of the change.%
    }
    \begin{center}
    \begin{footnotesize}
    \begin{tabular}{%
        @{}             l@{~~}r@{~~}r@{~~}r@{~~}r
        @{\hspace*{1em}}l@{~~}r@{~~}r@{~~}r@{~~}r@{}}
    \hline
    City
        & \multicolumn{1}{@{}c@{}}{Ave-All}
        & \multicolumn{1}{@{}c@{}}{Ave-19}
        & \multicolumn{1}{@{}c@{}}{L.R.}
        & \multicolumn{1}{@{}c@{}}{$R^2$} &
    City
        & \multicolumn{1}{@{}c@{}}{Ave-All}
        & \multicolumn{1}{@{}c@{}}{Ave-19}
        & \multicolumn{1}{@{}c@{}}{L.R.}
        & \multicolumn{1}{@{}c@{}}{$R^2$} \\
    \hline
Adana	&	75661.0	&	92369.2	&	0.51	&	0.96	&	K. Maras	&	28267.1	&	38666.2	&	0.30	&	0.99	\\
Adıyaman	&	15714.4	&	18490.8	&	0.09	&	0.94	&	Karabük	&	11220.9	&	14038.9	&	0.09	&	0.95	\\
Afyon	&	36326.0	&	48749.7	&	0.33	&	0.93	&	Karaman	&	10233.5	&	15320.2	&	0.14	&	0.97	\\
Ağrı	&	18604.6	&	27348.1	&	0.28	&	0.92	&	Kars	&	12523.4	&	16359.9	&	0.12	&	0.78	\\
Aksaray	&	13824.7	&	20837.6	&	0.19	&	0.91	&	Kastamonu	&	15388.0	&	19453.0	&	0.12	&	0.90	\\
Amasya	&	15369.1	&	20886.4	&	0.17	&	0.97	&	Kayseri	&	68485.9	&	84816.9	&	0.60	&	0.92	\\
Ankara	&	262707.9	&	312326.8	&	1.60	&	0.91	&	Kilis	&	5973.6	&	7511.4	&	0.04	&	0.92	\\
Antalya	&	117113.9	&	147358.0	&	0.91	&	0.99	&	Kırıkkale	&	10659.7	&	15008.0	&	0.11	&	0.94	\\
Ardahan	&	7620.3	&	12112.6	&	0.15	&	0.95	&	Kırklareli	&	16969.1	&	22042.6	&	0.13	&	0.96	\\
Artvin	&	5517.9	&	7104.9	&	0.05	&	0.84	&	Kırşehir	&	11767.5	&	15746.1	&	0.11	&	0.90	\\
Aydın	&	43808.9	&	51746.7	&	0.28	&	0.95	&	Kocaeli	&	104777.1	&	122444.0	&	0.52	&	0.93	\\
Balıkesir	&	51165.4	&	62837.4	&	0.43	&	0.94	&	Konya	&	91805.9	&	124538.6	&	0.89	&	0.96	\\
Bartın	&	6737.0	&	9765.1	&	0.08	&	0.97	&	Kütahya	&	24657.0	&	32040.4	&	0.21	&	0.93	\\
Batman	&	16369.0	&	22646.5	&	0.20	&	0.91	&	Malatya	&	28910.2	&	35651.4	&	0.18	&	0.93	\\
Bayburt	&	6388.6	&	7614.3	&	0.07	&	0.65	&	Manisa	&	48519.9	&	63196.2	&	0.48	&	0.97	\\
Bilecik	&	13312.5	&	16823.2	&	0.11	&	0.89	&	Mardin	&	25684.5	&	35221.4	&	0.28	&	0.85	\\
Bingöl	&	7050.5	&	9763.3	&	0.06	&	0.85	&	Mersin	&	66947.3	&	87291.0	&	0.57	&	0.98	\\
Bitlis	&	13654.8	&	18505.9	&	0.15	&	0.85	&	Muğla	&	48314.4	&	56420.4	&	0.27	&	0.91	\\
Bolu	&	14158.9	&	15359.3	&	0.07	&	0.65	&	Muş	&	9802.8	&	15372.0	&	0.15	&	0.91	\\
Burdur	&	14018.3	&	18886.4	&	0.13	&	0.98	&	Nevşehir	&	18262.5	&	22723.1	&	0.17	&	0.90	\\
Bursa	&	102608.8	&	121942.3	&	0.67	&	0.93	&	Niğde	&	20178.5	&	25379.2	&	0.17	&	0.95	\\
Çanakkale	&	20093.5	&	25955.5	&	0.20	&	0.96	&	Ordu	&	15431.2	&	20707.3	&	0.15	&	0.93	\\
Çankırı	&	9221.6	&	13460.1	&	0.12	&	0.95	&	Osmaniye	&	13712.8	&	17825.3	&	0.14	&	0.98	\\
Çorum	&	21636.7	&	27685.4	&	0.19	&	0.91	&	Rize	&	8369.0	&	12799.1	&	0.11	&	0.93	\\
Denizli	&	53840.5	&	62375.7	&	0.31	&	0.82	&	Sakarya	&	33663.1	&	41319.0	&	0.22	&	0.93	\\
Diyarbakır	&	41004.7	&	55799.3	&	0.42	&	0.95	&	Samsun	&	48676.5	&	59663.0	&	0.38	&	0.90	\\
Düzce	&	15901.6	&	19319.1	&	0.09	&	0.80	&	Şanlıurfa	&	42984.3	&	55148.2	&	0.47	&	0.92	\\
Edirne	&	18198.1	&	23596.3	&	0.11	&	0.87	&	Siirt	&	8143.0	&	12162.9	&	0.12	&	0.93	\\
Elazığ	&	21352.2	&	27436.9	&	0.17	&	0.96	&	Sinop	&	7448.3	&	9236.9	&	0.06	&	0.97	\\
Erzincan	&	14115.8	&	17225.3	&	0.13	&	0.84	&	Sivas	&	12290.2	&	17093.6	&	0.13	&	0.79	\\
Erzurum	&	47536.9	&	64825.1	&	0.50	&	0.84	&	Şırnak	&	29471.3	&	35903.9	&	0.29	&	0.87	\\
Eskişehir	&	38450.8	&	50304.5	&	0.28	&	0.93	&	Tekirdağ	&	43323.0	&	61303.2	&	0.42	&	0.94	\\
Gaziantep	&	56220.0	&	71179.5	&	0.46	&	0.98	&	Tokat	&	21508.0	&	29875.7	&	0.23	&	0.98	\\
Giresun	&	9376.0	&	14240.2	&	0.12	&	0.94	&	Trabzon	&	15022.6	&	24818.1	&	0.23	&	0.92	\\
Gümüşhane	&	5190.2	&	6610.6	&	0.05	&	0.75	&	Tunceli	&	6316.4	&	6308.7	&	0.00	&	0.01	\\
Hakkari	&	7313.8	&	11643.6	&	0.12	&	0.92	&	Uşak	&	17681.7	&	23676.8	&	0.16	&	0.99	\\
Hatay	&	57326.0	&	74518.1	&	0.49	&	0.98	&	Van	&	39477.1	&	58893.6	&	0.49	&	0.90	\\
Iğdır	&	7048.8	&	11293.2	&	0.12	&	0.90	&	Yalova	&	9416.6	&	11586.2	&	0.08	&	0.92	\\
Isparta	&	22095.5	&	29978.3	&	0.21	&	0.98	&	Yozgat	&	16876.3	&	24698.2	&	0.23	&	0.97	\\
İstanbul	&	368785.7	&	458720.4	&	1.92	&	0.88	&	Zonguldak	&	21851.1	&	28349.2	&	0.17	&	0.95	\\
İzmir	&	169984.4	&	202726.1	&	1.01	&	0.99	&		&		&		&		&		\\
    \hline
    \end{tabular}
    \end{footnotesize}
    \end{center}
\label{T:result}
\end{table*}
\begin{table*}
\caption{%
Turkish observatories listed in alphabetical order with their acronyms.
Their geographical locations and elevations were taken from \citep{aksaker2015}.%
}
\centering
\begin{tabular}{@{}lcc ccr@{}}
\hline 
\multicolumn{2}{@{}l}{Observatory Acronym \& Observatory Organisation}
    & City
    & $\lambda$
    & $\phi$
    & Elevation \\
\hline
AÜKR        & Ankara University Kreiken Observatory     & Ankara    &	32.78	&	39.84   &   1254 \\
ÇAAM        & Çanakkale Astrophysics Research Center    & Çanakkale &	26.48	&	40.01   &   373  \\
DAG         & Doğu Anadolu Gözlemevi                    & Erzurum 	&	41.23   &   39.78   &   3102 \\
EGE         & Ege University Observatory                & İzmir	    &	27.27	&	38.40   &   622  \\
İNÜ         & İnönü University Observatory              & Malatya   &	38.44	&	38.32   &   1021 \\
İÜO         & İstanbul University Göz. Uyg. Arş.Mrk.    & İstanbul 	&	28.96   &   41.01   &   55 \\
TUG         & TÜBİTAK National Observatory              & Antalya   &	30.34	&	36.82   &   2436 \\
TURAG       & Turkish National Radio Astronomy Obs. Site     & Karaman   &	33.09 	&	37.14     &   1062 \\
UZAYBİMER   & Astr. ve Uzay Bil. Göz. Uyg. Arş.Mrk.     & Kayseri   &	35.55	&	38.71   &   1094 \\
UZAYMER     & Uzay Bil. ve Güneş¸ En. Arş. Uyg. Mrk.    & Adana     &	35.35	&	37.06   &   112  \\
\hline 
\end{tabular}
\label{T:obs}
\end{table*}
\begin{table*}
\caption{%
Yearly average of AL for all observatories in Turkey. 
Column definitions are the same as in Table \ref{T:result}.
The SQM is in mpsas units, and SQM values were converted from AL for 2019.%
}
\begin{center}
\begin{footnotesize}
\begin{tabular}{%
        @{}@{~~}@{~~}r@{~~}r@{~~}r@{~~}r@{~~}c@{~~}c@{~~}r}
    \hline
    Obs.
    & Ave-All
    & Ave-19
    & L.R.
    & {$R^2$}
    & SQM
        \\
    \hline
AÜKR	    &	11.09	&	14.40	&	0.79	&	0.42	&	17.8	\\
ÇAAM	    &	0.26	&	0.30	&	0.01	&	0.48	&	21.0	\\
DAG 	    &	0.27	&	0.27	&	0.02	&	0.50	&	21.1	\\
EGE 	    &	1.79	&	2.21	&	0.14	&	0.80	&	19.3	\\
İNÜ 	    &	4.24	&	5.00	&	0.58	&	0.57	&	18.7	\\
İÜO 	    &	138.03	&	130.04	&	-3.09	&	0.40	&	16.0	\\
TUG     	&	0.10	&	0.14	&	0.01	&	0.60	&	21.6	\\
TURAG   	&	0.10	&	0.13	&	0.01	&	0.90	&	21.7	\\
UZAYBİMER	&	47.99	&	58.89	&	2.84	&	0.94	&	16.6	\\
UZAYMER	    &	10.47	&	9.53	&	-0.05	&	0.01	&	18.1	\\
    \hline
    \end{tabular}
    \end{footnotesize}
    \end{center}
\label{T:obs_result}
\end{table*}
\end{document}